\documentclass{PoS}

\usepackage{graphicx}

\newcommand{\msbar}{\overline{\rm MS}}
\newcommand{\preprintline}{\newline
\vskip -4.25cm
\rightline{\parbox{3.3cm}{\large\rm DESY 05-200 \\ Edinburgh 2005/15 \\
  LTH 669}}
\vspace{2.25cm}}

\PoS{PoS(LAT2005)125}

\title{Non-perturbative renormalisation for overlap fermions\preprintline}

\ShortTitle{Non-perturbative renormalisation for overlap fermions}

\author{Martin G\"urtler$^a$, Roger Horsley$^b$, Paul E. L. Rakow$^c$,
  \speaker{Charles J. Roberts}$\,^c$, Gerrit Schierholz$^{ad}$ and Thomas
  Streuer$^{ae}$\\
  \llap{$^a$}John von Neumann-Institut f\"ur Computing NIC, Deutsches
  Elektronen-Synchrotron DESY, 15738 Zeuthen, Germany\\
  \llap{$^b$}School of Physics, University of Edinburgh, Edinburgh EH9 3JZ,
  UK\\
  \llap{$^c$}Theoretical Physics Division, Department of Mathematical
  Sciences, University of Liverpool, Liverpool L69 3BX, UK\\
  \llap{$^d$}Deutsches Elektronen-Synchrotron DESY, 22603 Hamburg, Germany\\
  \llap{$^e$}Institut f\"ur Theoretische Physik, Freie Universit\"at
  Berlin, 14196 Berlin, Germany\\
  \llap{$^*$}E-mail: \email{cjr@liv.ac.uk}}

\author{QCDSF Collaboration}

\abstract{Using non-perturbative techniques we have found the renormalisation
  factor, $Z$, in the RI-MOM scheme for quark bilinear operators in quenched
  QCD. We worked with overlap fermions using the L\"uscher-Weisz gauge action.
  Our calculation was performed at $\beta=8.45$ at a lattice spacing of
  $1/a=2.1\,\mbox{GeV}$ using a value of $\rho=1.4$. Our results show good
  agreement between the vector and the axial vector in the zero mass limit.
  This shows that overlap fermions have good chiral properties. To attempt to
  improve the discretisation errors in our results we subtracted the
  $\mathcal{O}(a^2)$ terms in one-loop lattice perturbation theory from the
  Monte Carlo Green functions. In particular we paid attention to the
  operators for the observable $\langle x \rangle$. We found a value for the
  renormalisation constants $Z^{\msbar}_{v_{2b}}$ and $Z^{\msbar}_{v_{2a}}$
  just less than 1.9 at $\mu=1/a=2.1\,\mbox{GeV}$.}

\FullConference{XXIIIrd International Symposium on Lattice Field Theory\\
  25-30 July 2005\\
  Trinity College, Dublin, Ireland}

\begin{document}

  \section{Introduction}

  In a renormalisation scheme $\mathcal{S}$ a general renormalised operator
  is given by
  \begin{equation}
    \mathcal{O}^\mathcal{S}(M) = Z^\mathcal{S}_\mathcal{O}(M)
    \mathcal{O}^{bare},
  \end{equation}
  where $M$ is the scale. We have followed the Rome-Southampton
  method~\cite{Rome-Southampton} for determining the renormalisation constants
  of various lattice operators non-perturbatively in the regularisation
  invariant (RI-MOM) scheme. In particular we have paid attention to the
  operators for the observable $\langle x \rangle$.

  We have calculated the amputated Green functions, $\Lambda_\mathcal{O}$, for
  such operators with overlap fermions using the tadpole improved
  L\"uscher-Weisz gauge action in the quenched approximation. Our calculation
  was performed at $\beta=8.45$ at a lattice spacing of $1/a=2.1\,\mbox{GeV}$
  using a value of $\rho=1.4$. These are the same parameters that we used
  in~\cite{Z_V}. The gauge was fixed to the Landau gauge. We used a momentum
  source which results in smaller errors~\cite{QCDSF}.

  In both the RI-MOM and $\mbox{RI}^\prime$-MOM schemes we require
  \begin{equation}
    {Z^\mathcal{S}_\mathcal{O} \over Z^\mathcal{S}_\psi}\left.
    \Lambda_\mathcal{O}^{bare}\right|_{p^2=\mu^2}
    = \Lambda_\mathcal{O}^{tree} + \mbox{ other Dirac structures},
  \end{equation}
  where $Z_\psi^\mathcal{S}$ is the field renormalisation constant. So by
  measuring $\Lambda_\mathcal{O}^{bare}$ on the lattice we can find
  $Z^\mathcal{S}_\mathcal{O}/Z_\psi^\mathcal{S}$, but we need to eliminate
  $Z^\mathcal{S}_\psi$.

  The two schemes differ in their definition of $Z_\psi^\mathcal{S}$. In the
  $\mbox{RI}^\prime$-MOM scheme
  \begin{equation}
    \left. Z_\psi^{\rm{RI}^\prime}S\right|_{p^2=\mu^2} = S^{tree},
  \end{equation}
  where $S$ is the Landau gauge propagator. In the RI-MOM scheme
  \begin{equation}
    {1 \over Z_\psi^{\rm RI}}{1 \over 16N_C}\left. \sum_\nu Tr[\gamma_\nu
      \Lambda_{V_\nu^C}^{bare}]\right|_{p^2=\mu^2} = 1,
  \end{equation}
  where $V^C$ is the conserved vector current. Both present difficulties for
  overlap fermions. The propagator has large $\mathcal{O}(a^2)$ artefacts,
  which cannot be easily eliminate, and $V^C$ is hard to define or measure for
  overlap fermions.

  Fortunately, we know that the conserved vector current,
  $V^C$, is proportional to the local vector current, $V$, so we can find
  $Z_\psi^{\rm RI}$ from $\Lambda_V^{bare}$. Because it is conserved,
  $Z_{V^C}=1$. So
  \begin{equation}
    {1 \over Z^{\rm RI}_\psi}\Lambda_{V^C}^{bare} = {Z_V \over
      Z_\psi^{\rm RI}} \Lambda_V^{bare} \qquad \Leftrightarrow \qquad
    \Lambda_{V^C}^{bare} = Z_V\Lambda_V^{bare}.
  \end{equation}
  This allows us to measure $Z_V$ non-perturbatively by looking at the
  proton's charge and baryon number~\cite{Z_V}. At $\beta=8.45$, $Z_V=1.416$.

  So, if $Z_V$ is known, we can combine these relations to get
  \begin{equation}
    Z_\psi^{\rm RI} = Z_V{1 \over 16N_C} \left. \sum_\nu Tr[\gamma_\nu
      \Lambda_{V^C_\nu}^{bare}]\right|_{p^2=\mu^2},
  \end{equation}
  which allows us to find $Z_\mathcal{O}^{\rm RI}$ alone.

  The $v_{2b}$ and $v_{2a}$ operators are defined as
  $\overline{\psi}(\gamma_\mu T_{\mu\nu}D_\nu)\psi$, where $T_{\mu\nu}$ is a
  symmetric, traceless tensor. For the $v_{2b}$ operators $T_{\mu\nu}$ is
  diagonal. For the $v_{2a}$ operators it is off-diagonal. These operators are
  used to find $\langle x \rangle$ in a hadron. This observable is important
  in deep inelastic scattering.

  The running of the renormalised operator with the scale $M$ is controlled
  by the $\beta$ and $\gamma$ functions in the renormalisation group equation.
  These are defined as the scale derivatives of the renormalised coupling and
  the renormalisation constant:
  \begin{eqnarray}
    \beta^\mathcal{S}(g^\mathcal{S}(M)) &\equiv& \left. {\partial
      g^\mathcal{S}(M)\over\partial\log M} \right|_{bare},\\
    \gamma^\mathcal{S}_\mathcal{O}(g^\mathcal{S}(M)) &\equiv& \left.
    {\partial\log Z^\mathcal{S}_\mathcal{O}(M)\over\partial\log M}\right|
    _{bare},
  \end{eqnarray}
  where the bare parameters are held constant. These functions are
  given perturbatively as power series expansions in the coupling constant.
  Thus
  \begin{eqnarray}
    \beta^\mathcal{S}(g) &=& -b_0g^3-b_1g^5-b_2^\mathcal{S}g^7-b_3^\mathcal{S}
    g^9-\cdots,\\
    \gamma_\mathcal{O}^\mathcal{S}(g) &=& d_{\mathcal{O}0}g^2
    +d^\mathcal{S}_{\mathcal{O}1}g^4+d^\mathcal{S}_{\mathcal{O}2}g^6
    +d^\mathcal{S}_{\mathcal{O}3}g^8+\cdots.
  \end{eqnarray}
  The $\beta$ function is now known to four loops in the $\msbar$
  scheme~\cite{Chetyrkin}. For the scalar and pseudoscalar operators, the
  $\gamma$ function is known to four loops in the $\msbar$ and RI-MOM
  schemes~\cite{Chetyrkin}. For the tensor, pseudotensor, $v_{2b}$ and
  $v_{2a}$ operators it is known to three loops~\cite{Gracey-tensor,Gracey-x}
  in these schemes.

  We found the renormalisation group invariant (RGI) renormalisation constant
  using the relation
  \begin{equation}
    \mathcal{O}^{RGI} \equiv \Delta Z^\mathcal{S}_\mathcal{O}(M)O(M)
    = \Delta Z^\mathcal{S}_\mathcal{O}(M)Z^\mathcal{S}_\mathcal{O}(M)
    \mathcal{O}^{bare},
  \end{equation}
  where
  \begin{equation}
    [\Delta Z^\mathcal{S}_\mathcal{O}(M)]^{-1} =
    [2b_0g^{\msbar}(M)^2]^{-d_{\mathcal{O}0}/2b_0}
    \exp\left\{\int_0^{g^{\msbar}(M)}d\xi\left[
    {\gamma^\mathcal{S}_\mathcal{O}(\xi) \over \beta^{\msbar}(\xi)}
    + {d_{\mathcal{O}0} \over b_0\xi}\right]\right\}.
  \end{equation}
  We used anomalous dimensions in the RI-MOM scheme and a scale of $a^2p^2$.  We
  then converted the RGI renormalisation constant to the $\msbar$ scheme at
  the scale $a\mu=1$ using a similar relation involving the $\msbar$ anomalous
  dimensions.

  Even after this our data still contained lattice artefacts, especially at
  large $p^2$. To reduce these we calculated the same Green functions in
  lattice perturbation theory to one loop. We subtracted off the terms of
  leading order in the lattice spacing and regarded the remaining quantity of
  $\mathcal{O}(a^2)$ as a lattice artefact. This lattice artefact we
  subtracted from the Monte Carlo Green function and we compared the results
  before and after this subtraction. When finding the lattice perturbation
  theory result we used a coupling $g^{\msbar}(kp)$ and experimented with
  different values of $k$.

  \section{Results}

  \begin{figure}[htbp]
    \begin{center}
      \includegraphics[width=0.8\textwidth]{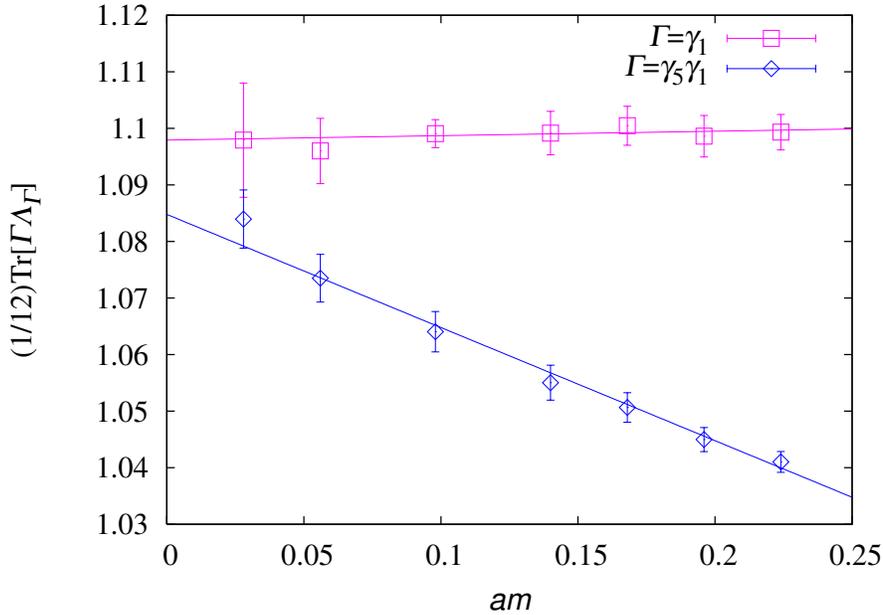}
      \caption{Here $am$ is the bare mass in lattice units. Our extrapolation
	to zero mass for a vector and an axial vector shows that there is good
	agreement between the two at zero mass for $a^2p^2=0.529$. At higher
	$p^2$ the agreement improves.}
      \label{fig:chiral_extrapolation}
    \end{center}
  \end{figure}
  \begin{figure}[htbp]
    \begin{center}
      \includegraphics[width=0.49\textwidth]{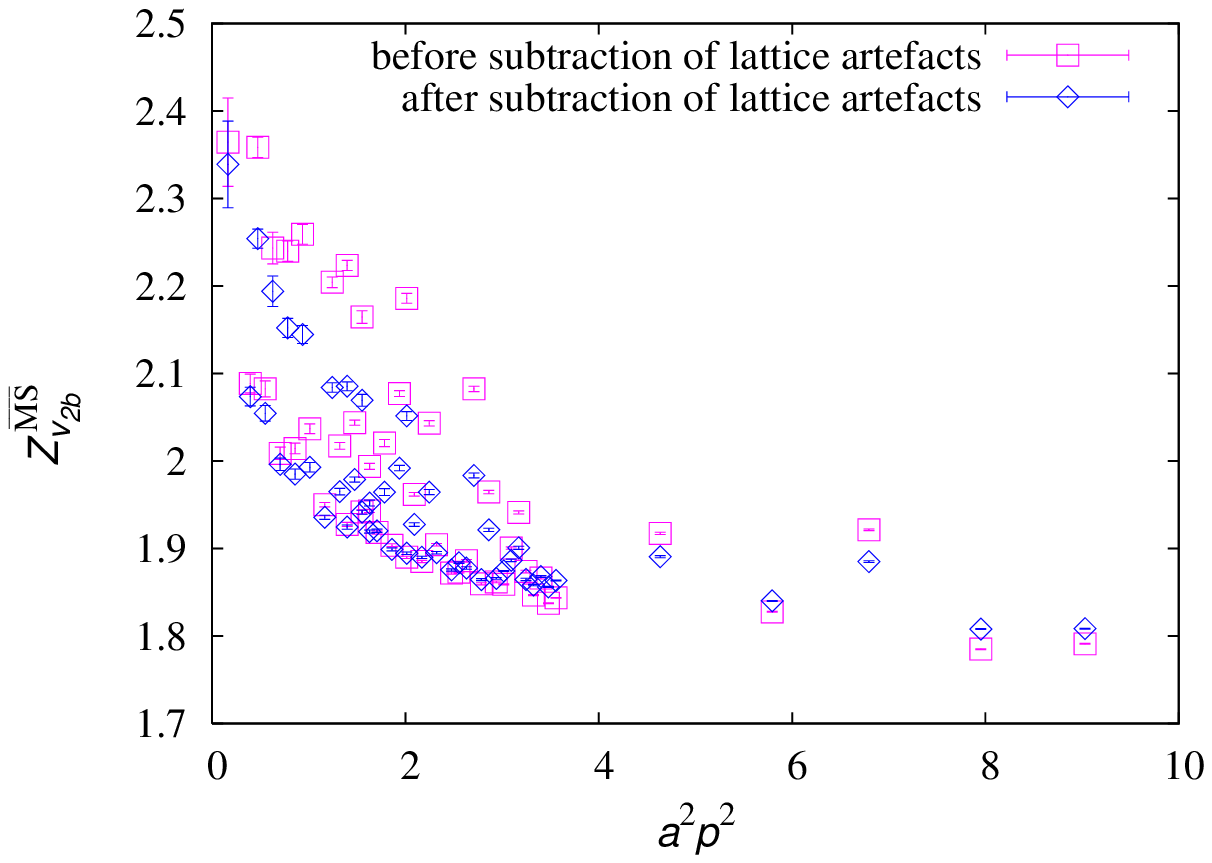}
      \includegraphics[width=0.49\textwidth]{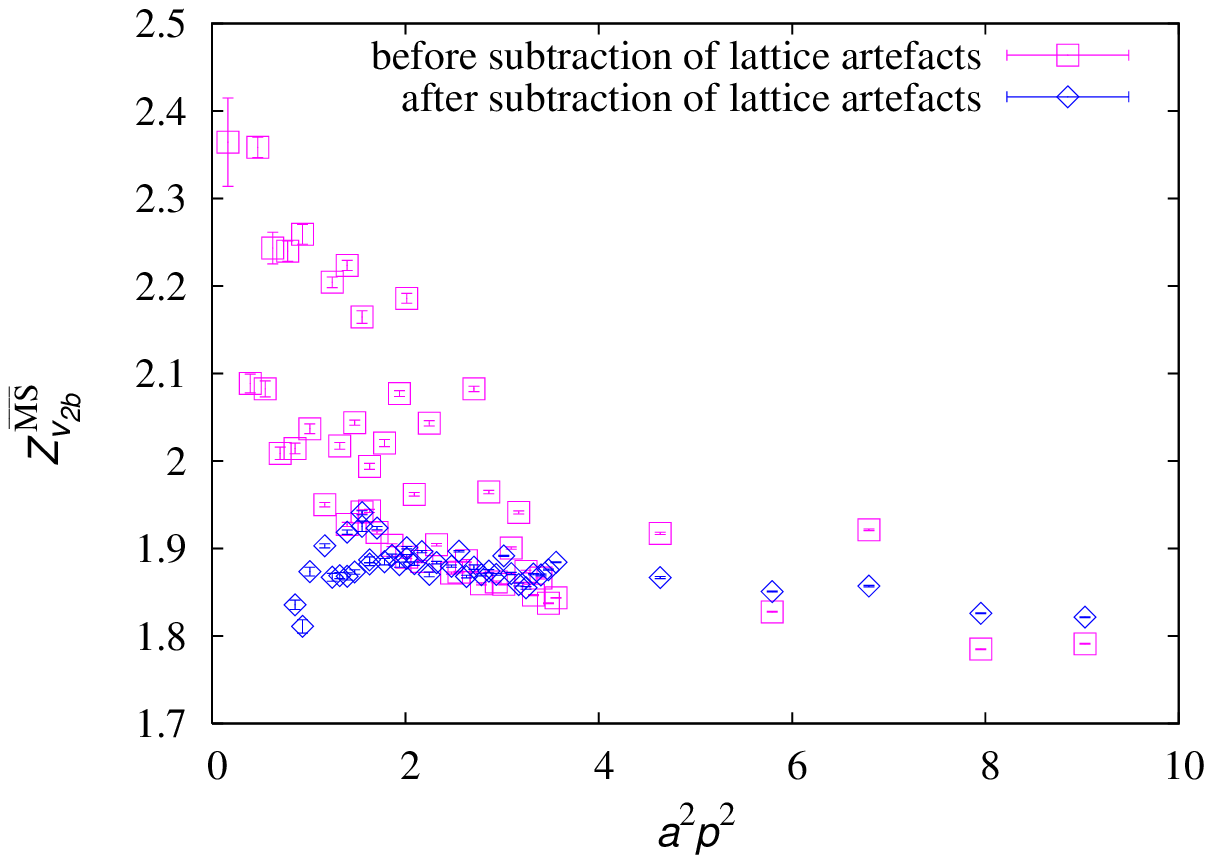}
      \caption{The operator shown here is $\overline{\psi}[\gamma_4D_4 -
	  1/3(\gamma_1D_1 + \gamma_2D_2 + \gamma_3D_3)]\psi$, where 4 is the 
	time direction. We obtained a value of the renormalisation constant
	$Z^{\msbar}_{v_{2b}}$ just less than 1.9 at $\mu=1/a=2.1\,\mbox{GeV}$.
	The coupling we used to calculate the artefacts on the left was
	$g^{\msbar}(p)$. The coupling we used to calculate the artefacts on
	the right was $g^{\msbar}(kp)$, where $k=0.22$. There are less
	points after subtraction of lattice artefacts since this coupling is
	not defined for the lower momenta.}
      \label{fig:z_v2b}
    \end{center}
  \end{figure}
  \begin{figure}[htbp]
    \begin{center}
      \includegraphics[width=0.8\textwidth]{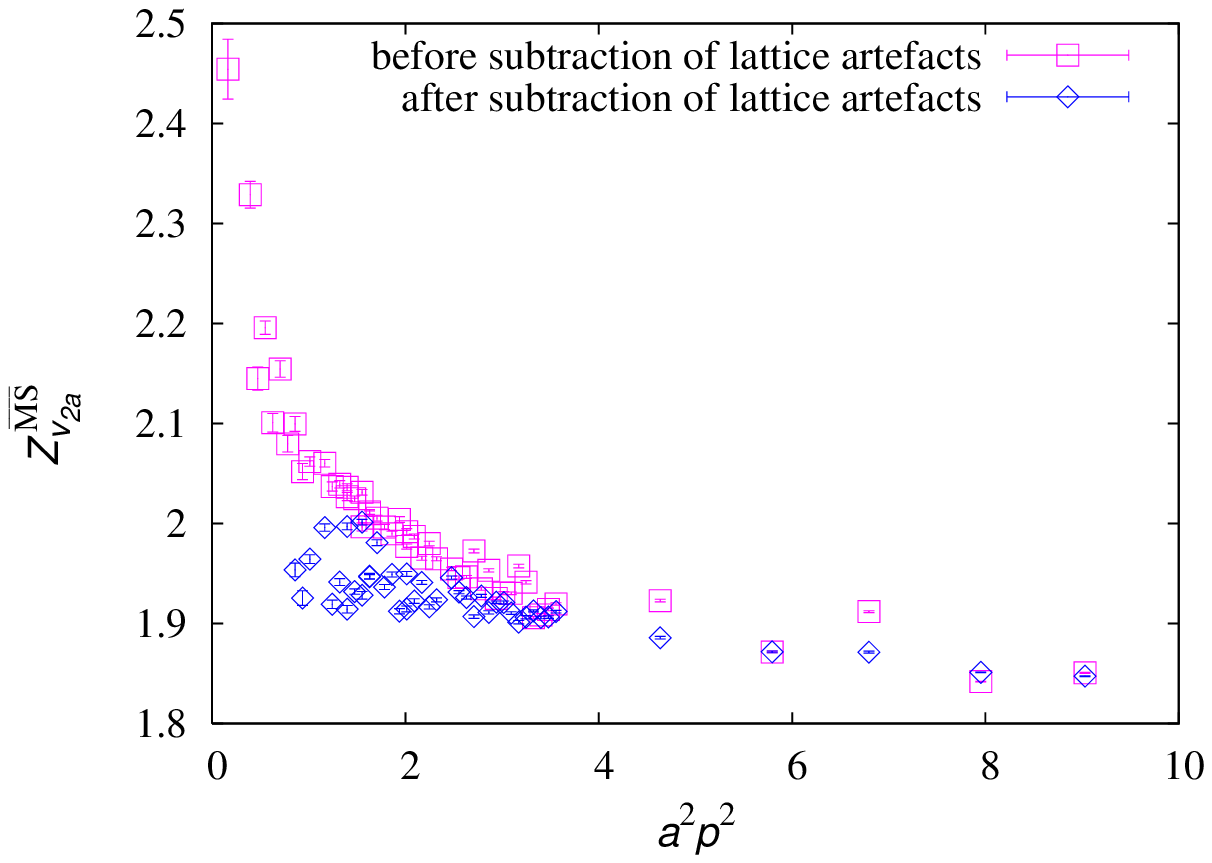}
      \caption{The operator shown here is $\overline{\psi}[1/\sqrt{2}
	(\gamma_1D_4 + \gamma_4D_1)]\psi$, where 4 is the time direction. We 
	obtained a value of the renormalisation constant $Z^{\msbar}_{v_{2a}}$
	just less than 1.9 at $\mu=1/a=2.1\,\mbox{GeV}$. The coupling we used
	to calculate the artefacts was $g^{\msbar}(kp)$, where
	$k=0.22$. There are less points after subtraction of lattice
	artefacts since this coupling is not defined for the lower momenta.}
      \label{fig:z_v2a_k0.05}
    \end{center}
  \end{figure}
  Firstly we tested to see whether the Green functions found using overlap
  fermions had better chiral properties than those found using conventional
  Wilson and clover fermions. Figure~\ref{fig:chiral_extrapolation} shows
  our extrapolation to zero mass for a vector and an axial vector at
  $a^2p^2=0.529$. At this momentum the agreement between the two results is
  quite good at zero mass. At higher values of the momentum the agreement
  improves further. This is not the case with Wilson and clover fermions and
  shows that overlap fermions have a proper chiral symmetry.

  Figure~\ref{fig:z_v2b} shows a plot of $Z^{\msbar}_{v_{2b}}$ against
  $a^2p^2$. There is a scale dependence because of the discretisation of the
  lattice. In particular, there is scatter, because the result depends on the
  directionality of the momentum as well as its magnitude. This is entirely an
  artefact of the lattice. We would expect this plot to be a perfect plateau
  because when we converted to the RGI $Z$ at a scale $a^2p^2$ this should
  have removed the scale dependence.
  
  To attempt to improve our plateau we calculated the $\mathcal{O}(a^2)$ terms
  for the same Green functions in one loop lattice perturbation theory and
  subtracted them from the Monte Carlo Green functions. On the left in
  figure~\ref{fig:z_v2b} we used a coupling $g^{\msbar}(p)$ for the
  lattice perturbation theory calculation. Our subtraction of the lattice
  artefacts gives some improvement in the right direction, but still not
  enough.

  It is possible to rescale the coupling to $g^{\msbar}(kp)$ in one loop
  lattice perturbation theory. This is because the difference between
  $g^{\msbar}(p)$ and $g^{\msbar}(kp)$ is $\mathcal{O}(g^4)$. It is not
  possible to tell the scale from one loop lattice perturbation theory alone.
  Thus we tried experimenting with rescaled couplings using different values
  of $k$ to see if this would improve our results.

  Plots for two of the operators we considered are shown in 
  figure~\ref{fig:z_v2b} on the right and figure~\ref{fig:z_v2a_k0.05}. These
  use a coupling $g^{\msbar}(kp)$, where $k=0.22$, in calculating the
  lattice artefacts. It is necessary to use a value of $k$ as small as this to
  significantly improve the plateau. Although at first sight the factor $k$
  might seem rather small, it is worth noting that the $\Lambda$ parameters
  which set the scale for different gauge actions can vary a great deal (see
  for example Table~6 of~\cite{scales}). So perhaps such a large rescaling is
  not unreasonable.
 
  Finally, from figure~\ref{fig:z_v2b} and figure~\ref{fig:z_v2a_k0.05},
  we obtained a value of the renormalisation constants $Z^{\msbar}_{v_{2b}}$
  and $Z^{\msbar}_{v_{2a}}$ just less than 1.9, which is in agreement
  with the results of~\cite{Thomas}.

  \section{Conclusion}

  The good agreement between the results for the vector and the axial vector
  in the zero mass limit shows that overlap fermions have a proper chiral
  symmetry. There would not be such an agreement with Wilson or clover
  fermions.

  We can improve our plateau in our results in figure~\ref{fig:z_v2b}
  and figure~\ref{fig:z_v2a_k0.05} quite significantly by choosing an
  appropriate value of $k$. We intend to investigate this further by doing a
  fit for the best value of $k$ and also by experimenting with different
  values of $\beta$ and seeing if the same value of $k$ also works there.

\end{document}